\documentclass[prb,reprint,amsmath,amssymb,showpacs,superscriptaddress,floatfix,longbibliography]{revtex4-1}
\usepackage[breaklinks=true,colorlinks,citecolor=blue,linkcolor=blue,urlcolor=blue]{hyperref}

\usepackage{epsfig,mathrsfs,color,latexsym,subfigure,marginnote,gensymb,}
\usepackage{graphicx}

\renewcommand{\BibitemShut}[1]{}

\begin{document}
\title{Strain-tunable charge carrier mobility of atomically thin phosphorus allotropes}

\author{Achintya Priydarshi}
\affiliation{Department of Electrical Engineering, Indian Institute of Technology, Kanpur, Kanpur 208016, India}
\author{Yogesh Singh Chauhan}
\affiliation{Department of Electrical Engineering, Indian Institute of Technology, Kanpur, Kanpur 208016, India}
\author{Somnath Bhowmick}
\email[]{bsomnath@iitk.ac.in}
\affiliation{Department of Materials Science and Engineering, Indian Institute of Technology, Kanpur, Kanpur 208016, India}
\author{Amit Agarwal}
\email[]{amitag@iitk.ac.in}
\affiliation{Department of Physics, Indian Institute of Technology, Kanpur, Kanpur 208016, India}

\date{\today}
\begin{abstract}
We explore the impact of strain on charge carrier mobility of monolayer $\alpha$, $\beta$, $\gamma$ and $\delta$-P, the four well known atomically thin allotropes of phosphorus, using density functional theory. Owing to the highly anisotropic band dispersion, the charge carrier mobility of the pristine allotropes is significantly higher (more than 5 times in some cases) in one of the principal directions (zigzag or armchair) as compared to the other. Uniaxial strain (upto 6\% compressive/tensile) leads to bandgap alteration in each of the allotropes, especially a direct to indirect bandgap semiconductor transition in $\gamma$-P and a complete closure of the bandgap in $\gamma$ and $\delta$-P. We find that the charge carrier mobility is enhanced typically by a factor of $\approx 5-10$ in all the allotropes due to uniaxial strain; notably among them a $\approx 250$ (30) times increase of the hole (electron) mobility along the armchair (zigzag) direction is observed in $\beta$-P ($\gamma$-P) under a compressive strain, acting in the armchair direction. Interestingly, the preferred electronic conduction direction can also be changed in case of $\alpha$ and $\gamma$-P, by applying strain. 

\end{abstract}
\maketitle
\section{INTRODUCTION}
\label{intro}
Since the first successful mechanical exfoliation of monolayer graphene,\cite{Novoselov2004} atomically thin two dimensional (2D) materials have attracted immense interest, both in terms of fundamental physics and potential device applications. As the low energy quasiparticles of graphene behave like massless Dirac fermions; in the absence of charge impurities and ripples, it has an astonishing intrinsic carrier mobility value of 200,000 cm$^2$V$^{-1}$s$^{-1}$ at room temperature, nearly three times higher than that of InSb, known to have the highest mobility among inorganic semiconductors.\cite{chen2008intrinsic,hrostowski1955} Although external factors reduce the mobility significantly, still very large mobility in the range of 3,700--15,000 cm$^2$V$^{-1}$s$^{-1}$ is routinely reported, irrespective of underlying substrate and growth mechanism.\cite{chen2008intrinsic,kim2009large} However, due to lack of bandgap, graphene based transistors are known to have very large off current, leading to high static power consumption and thus, not suitable for applications in low power logic devices.\cite{schwierz2010graphene} On the other hand, transition metal dichalcogenides (TDMs)\cite{wang,MOS2,SNSE2} are endowed with a direct bandgap; for example, measuring $\sim$1.8 eV for monolayer MoS$_2$.\cite{mak2010atomically} As a result, MoS$_2$ based transistors are reported to exhibit I$_{ON}$/I$_{OFF}$ as high as $10^8$ at room temperature,\cite{radisavljevic2011single} although the carrier mobility is found to be much less than that of graphene, typically of the order of 200 -- 700 cm$^2$V$^{-1}$s$^{-1}$ for n-type FET (field effect transistor) operation.~\cite{popov2012,das2013}

Recently, monolayer of black phosphorus or phosphorene ($\alpha$-P) has emerged as a promising semiconductor, with a direct bandgap of magnitude 1.5 eV and as a consequence, transistors based on phosphorene are reported to exhibit very good I$_{ON}$/I$_{OFF}$ ratio of $10^4-10^5$, in addition to a reasonably high carrier mobility of 1000 -- 2000 cm$^2$V$^{-1}$s$^{-1}$ at the room temperature.\cite{liu2014phosphorene,li2014electrons,kou2015phosphorene,li2014black,churchill2014two,wei2016} Several other stable allotropes of monolayer phosphorus have also been predicted so far. \cite{semicond.blue,nahas2017} 
Of all the possible allotropes of monolayer phosphorous, the four most stable ones (in decreasing order of stability) are named as $\alpha$, $\beta$, $\gamma$ and $\delta$-P, and these four are the subject of this paper. Of these four,  $\alpha$-P is the most stable due to very small interfacial energy.\cite{guan2014phase} The other three, 
namely $\beta$, $\gamma$ and $\delta$-P are interesting because these phases can be connected (to form an in plane heterostructure, such that more than one phase coexist within a layer) easily with each other.\cite{guan2014phase} Being atomically thin layers, electronic band structure of phosphorene allotropes can be tuned easily by physical or chemical adsorption of adatoms,\cite{rastogi,nahasprb,C4CP03890H} out of the plane electric field\cite{ghoshbp,PhysRevB.94.205426,PhysRevB.93.195428} and applied strain.\cite{sa2014strain,anisotropy} Similar studies have reported stress induced bandgap modification in Si nanowires\cite{lu2007} and SnSe.\cite{wu2017}

In this work, using \textit{ab initio} calculations, we investigate the effect of strain on the acoustic phonon limited charge carrier mobility of $\alpha$, $\beta$, $\gamma$ and $\delta$-P. 
A remarkable feature, which sets these phosphorus allotropes apart from other 2D materials like graphene, TMDs etc., is the anisotropy observed between the two principal axes, aligned parallel to the zigzag and armchair direction, respectively.\cite{nahas2017,guan2014phase,ghosh2017,0953-8984-29-28-285601,C5NR00355E} Because of their anisotropic nature, atomically thin phosphorus allotropes are expected to respond differently, depending on whether the external stimulus (strain in this work) is applied in the zigzag or armchair direction; as reported for bandgap tuning depending on the direction of applied strain.\cite{guan2014phase} Prompted by this, we apply uniaxial strain along the zigzag and armchair direction separately; and other than bandgap closure, our study also reveals significant increase of carrier mobility, as well as change of the preferred electronic conduction direction, which clearly shows the potential of $\alpha$, $\beta$, $\gamma$ and $\delta$-P for making strain controlled electronic devices.  

\begin{figure*}
\begin{center}
\includegraphics[width=0.99 \linewidth]{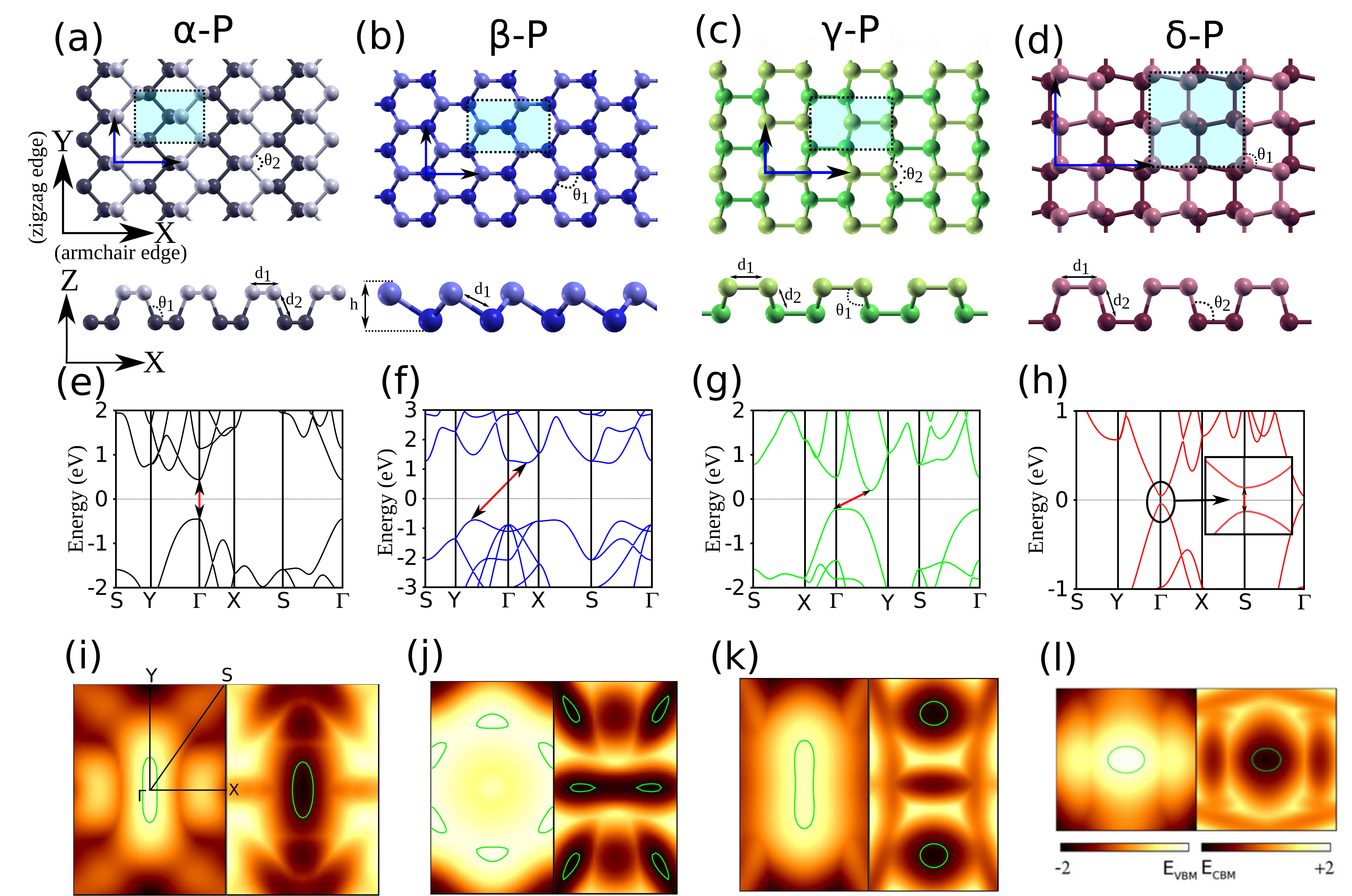}
\caption{ Panels (a), (b), (c) and (d) display the crystal structures (top and side view) of $\alpha$-P (monolayer of black phosphorus), $\beta$-P (monolayer of blue phosphorus), $\gamma$-P and $\delta$-P, respectively. Corresponding electronic band structures are shown in panels (e), (f), (g) and (h) and the 2D plots of the valence (left panel) and conduction band (right panel) energy over the entire Brillouin zone are illustrated in (i), (j), (k) and (l), where the panel sizes are according to the size of the Brillouin zone. Values of the lattice parameters and $d_1, d_2, h, \theta_1, \theta_2$ are reported in Table~\ref{t1}. Bright (dark) regions in the 2D plots [panels (i)-(l)] represent high (low) energy within a particular band. The green curve marks an energy contour line just below (above) the valence band maximum (conduction band minimum). All the energy contour lines are elliptic in shape, which clearly reflects the anisotropy of electrons and holes in terms of effective mass. Crystal structures are prepared by using Xcrysden.\cite{Kokalj03}
\label{f1}}
\end{center}
\end{figure*}

This paper is organized as follows. Computational details including 1) parameters for \textit{ab initio} calculations and 2) details of mobility calculation are presented in Sec.~\ref{cd}. This is followed by a brief description of crystal structure and electronic band structure of pristine 2D allotropes of phosphorus in Sec.~\ref{csebs}. Results related to strain dependent mobility calculations are presented and analyzed in Sec.~\ref{rd} and finally, we summarize our findings in Sec.~\ref{con}. 

\section{Computational details}
\label{cd}
\subsection{ab-initio calculations}
Crystal structure optimizations and electronic band structure calculations are performed within the framework of density functional theory (DFT), as implemented in the QUANTUM ESPRESSO\cite{QE} package, using a plane-wave basis set (kinetic energy cutoff = 30 Ry) and projector augmented-wave (PAW) pseudopotential. The  exchange and correlation  effects are treated using generalized gradient approximation (GGA).\cite{PBE} Two dimensionality is ensured by incorporating a vacuum space of 1.5 nm, perpendicular to the plane of the atoms, such that out of plane interaction between adjacent layers can be neglected. For $\alpha-$phosphorene, a $k$-point grid of $24\times 32\times 1$ is used for the 2D Brillouin zone integrations and the grid size is appropriately scaled for other allotropes, according to their unit cell dimensions. Structural relaxations are carried out until the forces on each atom are less than $10^{-3}$ Ry/au.

\subsection{Calculation of electron and hole mobility}
\label{cmob}
Charge carriers are mainly scattered by defects and phonons, which limits the mobility of electrons/holes in a semiconductor.   The latter is also primarily responsible for the temperature dependence of the electrical conductivity. In this work, we focus on pristine materials without any defect and thus, calculate intrinsic phonon limited mobility at the room temperature. Using deformation potential theory and effective mass approximation, the phonon limited mobility of charge carriers in a 2D crystal, along the $\alpha$ ($\alpha=x $/$y$) direction can be expressed as,\cite{shui,takagi,bruzzone2011ab,shao2013first,qiao2014high,peng2014strain}
\begin{equation}
\label{eqm}
\mu _{\alpha}=\frac{e\hbar^3}{\lambda_\alpha m^*_{\alpha}\sqrt{{m^*_\alpha}{m^*_\beta}} },
\end{equation}
where $m^*_{\alpha}$ and $m^*_{\beta}$ is the carrier effective mass in the transport and it's transverse direction, respectively 
and $\lambda_\alpha$ is the scattering probability for the charge carriers, given by
\begin{equation}
\label{eql}
\lambda_\alpha=\frac{k_BT(E_\alpha^i)^2}{C_\alpha}. 
\end{equation}
Here  $C_{\alpha}$ is the elastic constant in the direction of propagation of the longitudinal acoustic (LA) phonon (aligned along the $\alpha$ direction, which is also parallel to the transport direction); $T$ is the temperature;   and  $E^i_\alpha$ is the deformation potential. 

The elastic modulus along the $\alpha$  direction ($C_{\alpha}$) is defined via the relation $(E-E_0)/S_0 = \frac{1}{2} C_\alpha (\Delta{l}_\alpha/l_{\alpha 0})^2$, where $S_0$ is the area of the unstrained unit cell and $(E-E_0)$ is the change in total energy after applying a strain of $\Delta{l}_\alpha/l_{\alpha 0}$. Value of $C_\alpha$ is obtained via parabolic fitting of strain vs. energy curve, where the strain is generally varied from -2\% to 2\%. The effective mass is defined as $m^{-1}_{\alpha}\equiv m^{-1}_{\alpha\alpha}=\frac{1}{\hbar^2}\frac{\partial ^2E}{\partial k_\alpha\partial k_\alpha}$, and is obtained by parabolic fitting of $E$ vs. $k$ curve for electrons and holes, in vicinity of the VBM and CBM, respectively. The deformation potential is defined as $E^i_{\alpha}=\Delta{E_i}/(\Delta{l_\alpha/l_{\alpha 0}})$, where $\Delta{E_i}$ is the energy change of the $i^{th}$ band under applied strain of $\Delta{l_\alpha/l_{\alpha 0}}$. The energy change of the $i^{th}$ band is calculated with respect to the potential in the vacuum. Value of $E^i_{\alpha}$ is obtained from a linear fit of $\Delta{E_i}$ vs. strain, where the strain is gradually varied from -2\% to 2\% and $i$ is taken to be the VBM and CBM for holes and electrons, respectively. 

Our paper primarily focusses on the acoustic phonon limited mobility, which generally dominates the transport for small bias voltages. However, 
at moderate to high fields, transport is also dominated by scattering due to the optical phonons \cite{doi:10.1063/1.4960526} and other scattering mechanisms.  
We also emphasize that in case of acoustic phonon limited mobility calculation,  the deformation potential calculation is very sensitive to the choice of DFT parameters etc. This results in some quantitative discrepancy in the mobility calculations as can be seen from comparison of our results with Refs.~[\onlinecite{qiao2014high,10.1038/srep09961}]. 
However, acoustic phonon limited mobility based on DFT calculations still captures the qualitative trends in the mobility for different structures and for each structure with strain.


\subsection{Crystal structure and electronic band structure}
\label{csebs}
\subsubsection{Pristine allotropes}
The crystal structures of all four allotropes of monolayer phosphorus are shown in Fig.~\ref{f1} (a)-(d). Because of the sp$^3$ hybridization, phosphorus monolayers are not flat and have either puckered or buckled structure. Optimized structural parameters are shown in Table~\ref{t1} and they are in good agreement (within 5\%) with the values reported in the literature.\cite{guan2014phase,semicond.blue,nahas2017} The electronic band structures of the pristine allotropes are shown in  Fig.~\ref{f1} (e)-(h); while two of the allotropes, namely $\alpha$ and $\delta$ are identified to be the direct bandgap semiconductors, rest of them ($\beta$ and $\gamma$) are found to be indirect in nature. Comparing our results with those reported in the literature,\cite{guan2014phase,semicond.blue,nahas2017} both qualitative and quantitative agreement is observed, thus validating the simulation parameters described in Sec.~\ref{cd}. We also illustrate the 2D plot of valence and conduction band energies over the entire Brillouin zone for the four undeformed allotropes in Fig.~\ref{f1}(i)-(l). The green line marks a constant energy contour line just below the valence band maximum (VBM) and just above the conduction band minimum (CBM) energy. All the constant energy contour lines are elliptical in shape, which clearly reflects the anisotropy of electrons and holes in terms of effective mass.

\begin{table}
\caption{Equilibrium structural parameters of all four monolayer allotropes of phosphorus, where $d_1, d_2$ are the bond lengths, $h$ is the thickness of the layer, $\theta_1, \theta_2$ are the bond angles [see Fig~\ref{f1}(a)-(d)].\label{t1}}
\centering
\begin{tabular}{c c c c c c}
\hline
Phase:    & $\alpha$-P & $\beta$-P & $\gamma$-P & $\delta$-P \\
\hline
$d_1,d_2(\AA)$  & 2.22, 2.25 & 2.26 & 2.24, 2.32 & 2.23, 2.27\\
h (\AA)  & 2.11 & 1.23 & 1.49 & 2.14 \\
$\theta_1, \theta_2 (\degree)$ & 103.81, 96.05 & 93.09 & 100.02, 93.34 & 101.20, 101.07 \\
$|\bf a_x| $ (\AA)  & 4.58 & 5.69 & 5.43 & 5.53 \\
$|\bf a_y|$ (\AA)  & 3.30 & 3.28 & 3.27 & 5.41 \\
\hline
\end{tabular}
\end{table}

\subsubsection{Strained allotropes}
After successfully reproducing the structural and electronic properties of pristine monolayer phosphorus allotropes, we  now explore the effect of strain on the electronic band structure.
Uniaxial strain is applied separately along the two mutually perpendicular directions, parallel to the $x$ (armchair) and $y$ (zigzag) axis [see Fig.~\ref{f1}]. Both compressive (negative) and tensile (positive) strain is applied, while the magnitude of the strain is varied upto 6 $\%$. Since 2D materials have superior mechanical flexibility, like $\alpha$-P can withstand approximately up to 30\% strain,\cite{peng2014} we believe that the allotropes remain stable over the range of deformations considered in this paper. The applied strains $\sigma_x$ and $\sigma_y$ are defined as $\sigma_x=(a_x-a_{x0})/a_{x0}, \sigma_y=(a_y-a_{y0})/a_{y0}$, where $a_x, a_y$ and $a_{x0}, a_{y0}$ are the lattice parameters of the strained and pristine unit cells along the $x$ and $y$-direction, respectively.

\begin{figure}
\includegraphics[width=\linewidth]{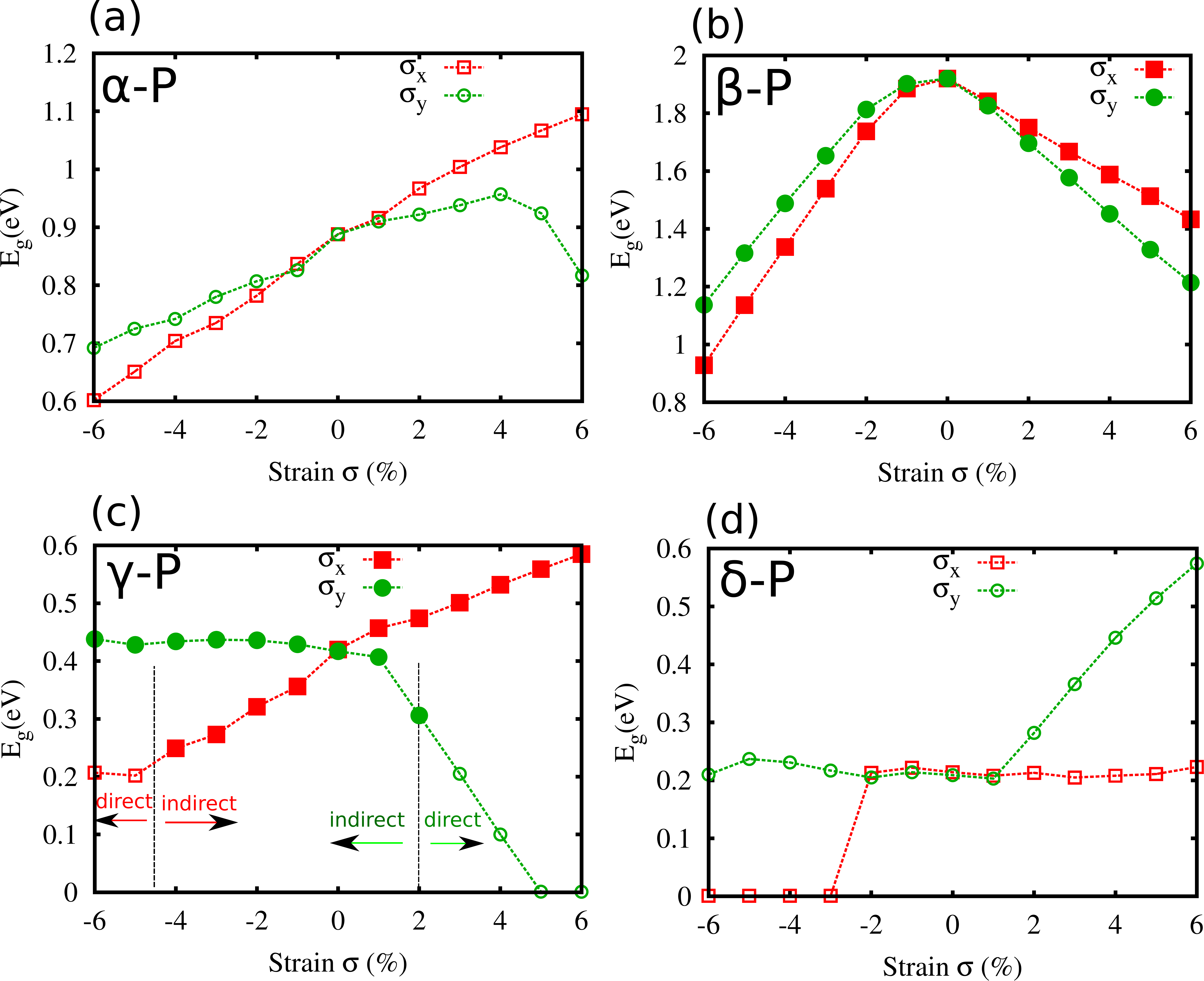}
\caption{Panels (a), (b), (c) and (d) display the change of electronic bandgap of monolayer phosphorus allotropes as a function of uniaxial strain along the $x$ and $y$ axis. The filled and hollow circles indicate an indirect and a direct bandgap semiconductor, respectively. Indirect to direct bandgap transition is observed only in case of monolayer $\gamma$-P, at $\sigma_x\le -5\%$ and $\sigma_y\ge 3\%$. Bandgap reduces to zero at $\sigma_y\ge 5\%$ and  $\sigma_x\le -3\%$, in $\gamma$-P and $\delta$-P, respectively.}
\label{f2}
\end{figure}

Electronic band structure of phosphorene allotropes are known to be significantly modified by applied strain.\cite{guan2014phase,semicond.blue} Similar outcomes are obtained in our calculations, which include bandgap enhancement/closure and a transition from indirect to direct bandgap semiconductor. As shown in Fig.~\ref{f2}, most dramatic effects are observed in case of monolayer $\gamma$ and $\delta$-P under applied strain. For example, pristine $\gamma$-P, which is an indirect bandgap semiconductor, is converted to a direct bandgap semiconductor at a modest strain of $\sigma_y=3\%$ (also observed at $\sigma_x=-5\%$), followed by a complete closure of the bandgap at $\sigma_y=5\%$. Similarly, for $\delta$-P bandgap diminishes to zero at $\sigma_x=-3\%$. Although no such transition is observed in case of monolayer $\alpha$ and $\beta$-P, however, the magnitude of the bandgap varies $\pm (30-40)\%$ from their pristine values when they are subjected to external strain. Note that, since DFT calculations underestimate the bandgap, complete bandgap closure might happen at higher strain than reported here. Nevertheless, electron and hole mobility are likely to follow the calculated trend since DFT captures the structural properties well, and the bandgap magnitude hardly plays any role in carrier mobility.  Owing to their close connection to the electronic band structure, electron and hole mobilities are also expected to be modified via applied strain, and the strain dependent mobility is discussed in detail in the rest of the paper. 

\begin{figure}
\includegraphics[width=\linewidth]{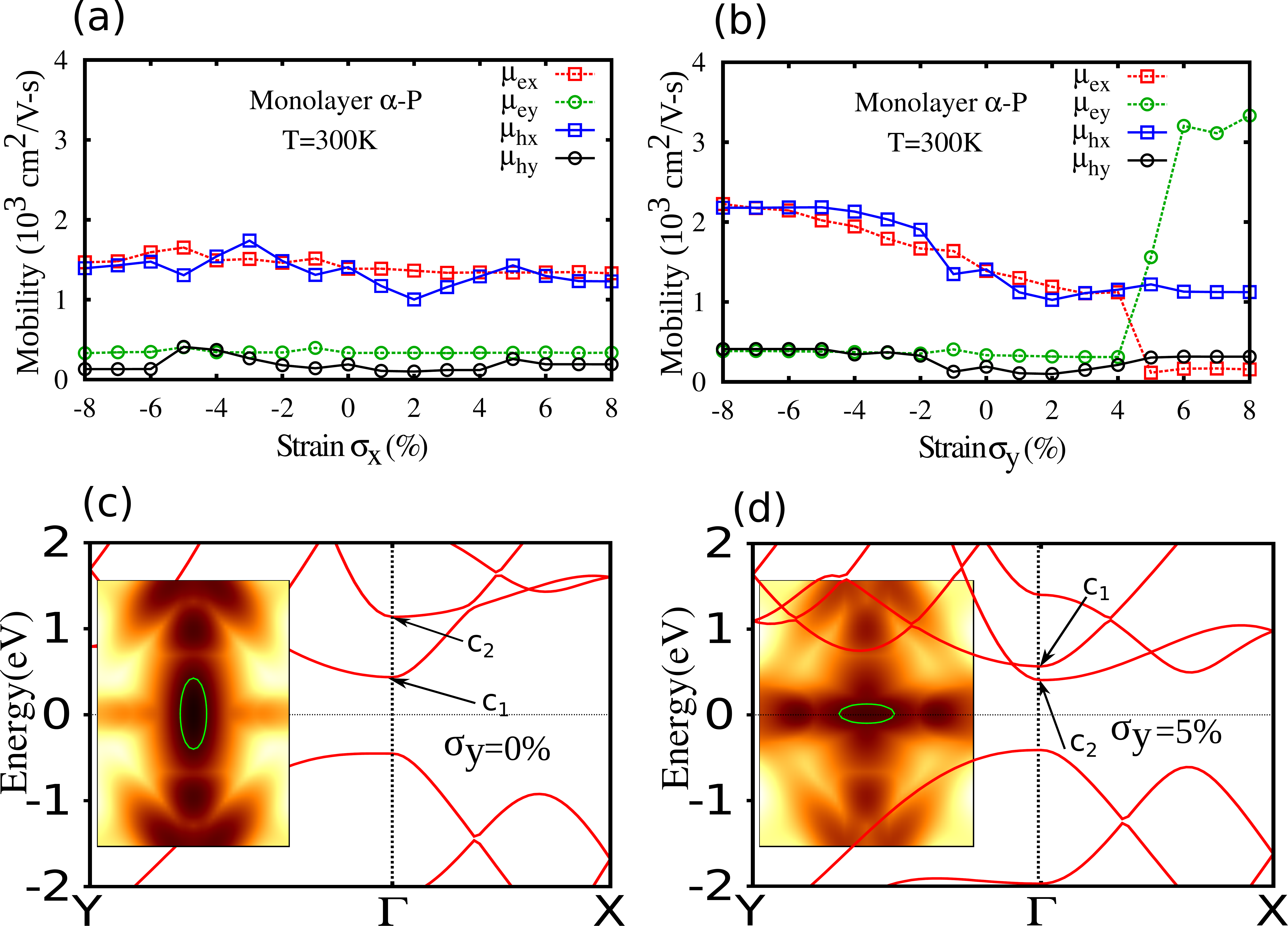}
\caption{Panels (a), (b) display the strain dependence of electron and hole mobility of $\alpha$-P. The most significant effect is observed at 5\% tensile strain ($\sigma_y$), when preferred electron conduction direction is switched from armchair ($x$ axis) to zigzag ($y$ axis). Comparing the electronic band structure of pristine [panel (c)] and strained [panel (d)] $\alpha$-P, it is observed that the CBM changes due to strain, leading to reversal of the direction along which effective mass of electron has the highest and the lowest value, which explains the sudden change of electron mobility at $\sigma_y=5\%$. The insets in (c) and (d) show the 2D plots of the conduction band energy over the entire Brillouin zone, for pristine and strained $\alpha$-P.}
\label{f3}
\end{figure}

\begin{table*}[t]
\caption{Carrier mobility and other relevant parameters of pristine $\alpha$, $\beta$, $\gamma$ and $\delta$-P. We take the temperature as 300K.} 
\label{t2}
\centering
\begin{tabular}{c c c c c c c c c c c c}
\hline \hline
Phase	& Carrier & $m^{*}_{x}/m_{0}$ & $m^{*}_{y}/m_{0}$& $E_{x}^{c/v}$  & $E_{y}^{c/v}$ &$C_{x}$ &$C_{y}$ & $\lambda_x$ &$\lambda_y$ &$\mu_{x}$ &$\mu_{y}$  \\
& type & & &(eV) & (eV) & (Jm$^{-2}$) & (Jm$^{-2}$) & (eV$^2$\AA$^2$) & (eV$^2$\AA$^2$) & (10$^3$cm$^2$V$^{-1}$s$^{-1}$) & (10$^3$cm$^2$V$^{-1}$s$^{-1}$)\\
\hline \hline
$\alpha$-P  & e & 0.145 & 1.262 & 2.59 & 3.24 & 27.73 & 90.53 & 0.1008 & 0.0483 & 1.38 & 0.333  \\
 & h & 0.138 & 3.684 & 2.05 & 1.95 & 27.73 & 90.53 & 0.0631 & 0.0175 & 1.40 & 0.189  \\
& & \\
 $\beta$-P  & e & 0.945 & 0.137 & 3.32 & 7.94 & 76.37 & 77.39 & 0.0601 & 0.3393 & 0.426 & 0.522  \\
 & h &  3.840 & 0.8534 & 0.54 & 1.19 & 76.37 & 77.39 & 0.0016 & 0.0077 & 0.788 & 0.728  \\
 & & \\
$\gamma$-P  & e & 0.363 & 0.480  & 1.48 & 2.20 & 82.73 & 68.92 &  0.0110 & 0.0293 & 5.22 & 1.49 \\
& h & 0.398 & 7.520 & 6.20 &  2.56 & 82.73 & 68.92 & 0.1935 & 0.0396 & 0.065 & 0.017 \\
& & \\
$\delta$-P  & e & 0.387 & 0.169 & 2.98 & 2.26 & 47.35 & 81.66 & 0.0781 & 0.0261 &1.13 & 7.70 \\
& h & 0.412 & 0.146 & 2.94 & 2.42 & 47.35 & 81.66 & 0.0760 & 0.0299 & 1.14 & 8.18 \\
\hline
\end{tabular}
\end{table*}

\section{Analysis of Mobility}
\label{rd}
In this section, we discuss the charge carrier mobilities of monolayer phosphorene allotropes, calculated using the methodology described in Sec.~\ref{cmob}. Required parameters [in Eq.~\ref{eqm} and Eq.~\ref{eql}] like effective mass, deformation potential and elastic constant are obtained from the \textit{ab initio} calculations, as detailed in Sec.~\ref{cd}. The temperature is set to room temperature, i.e., 300 K. 

\subsection{Mobility of monolayer $\alpha$-P}
Monolayer phosphorene or $\alpha$-P is a direct bandgap semiconductor, with a unique electronic band structure, having a ``Schr\"odinger like'' quadratic band dispersion along the $\Gamma-Y$ (zigzag) direction, which is in stark contrast to it's ``massive Dirac like'' band dispersion along the $\Gamma-X$ (armchair) direction. Accordingly, the effective mass of electron (hole) is found to be 8.7 (26.7) times larger in the zigzag direction, as compared to the armchair direction [see Table~\ref{t2}]. Among the other parameters controlling the charge carrier mobility, the elastic constant in the zigzag direction is 3.25 times larger than it's value in the armchair direction, while the anisotropy of the deformation potential is relatively moderate [see Table~\ref{t2}]. Accordingly, the scattering probability of electron (hole) is 2 (3.6) times higher in the armchair direction, than that of in the zigzag direction. However, effective mass anisotropy (being much larger than $\lambda$ anisotropy) dominates and the electron (hole) mobility at room temperature is 4.1 (7.4) times higher along the armchair direction, as compared to the zigzag direction. 

Further extending the study to strained $\alpha$-P, we find that $\sigma_x$ does not have any significant effect on charge carrier mobility [see Fig.~\ref{f3}(a)]. However, applied strain along the zigzag direction ($\sigma_y$) changes the electron mobility (in both directions), as well as the hole mobility (only in the $x$ direction), substantially. The most prominent difference is observed at 5\% tensile strain, which alters the preferred direction of conduction from armchair (in pristine $\alpha$-P) to zigzag. As shown in Fig.~\ref{f3}(b), there is approximately 10 times increase of electron mobility along the zigzag  direction at 5\% and higher tensile strain, as compared to it's value in pristine $\alpha$-P. This can be understood by comparing the electronic band structure of pristine and strained $\alpha$-P [see Figs.~\ref{f3}(c)-(d)]. Focusing on the first two bands (marked as $c_1$ and $c_2$ in the figure) above the Fermi level, while $c_1$ is the CBM in case of pristine $\alpha$-P, but $c_2$ occupies this position when subjected to 5\% (or more) tensile strain along the $y$ direction. This reverses the anisotropy, with the effective mass now being heavier along the armchair axis than that of zigzag axis and thus the latter becomes the preferred electron conduction direction for tensile strain  $\sigma_y \ge 5\%$. The change in the effective mass anisotropy can also be inferred from the 2D plots of conduction band energy over the entire Brillouin zone, shown in the inset of Fig.~\ref{f3}(c) (pristine) and Fig.~\ref{f3}(d) ($\sigma_y=5\%$). Marginal increase of hole mobility from it's pristine value is also observed, when $\sigma_y$ is compressive [see Fig.~\ref{f3}(b)].

\begin{figure}[b]
\includegraphics[width=\linewidth]{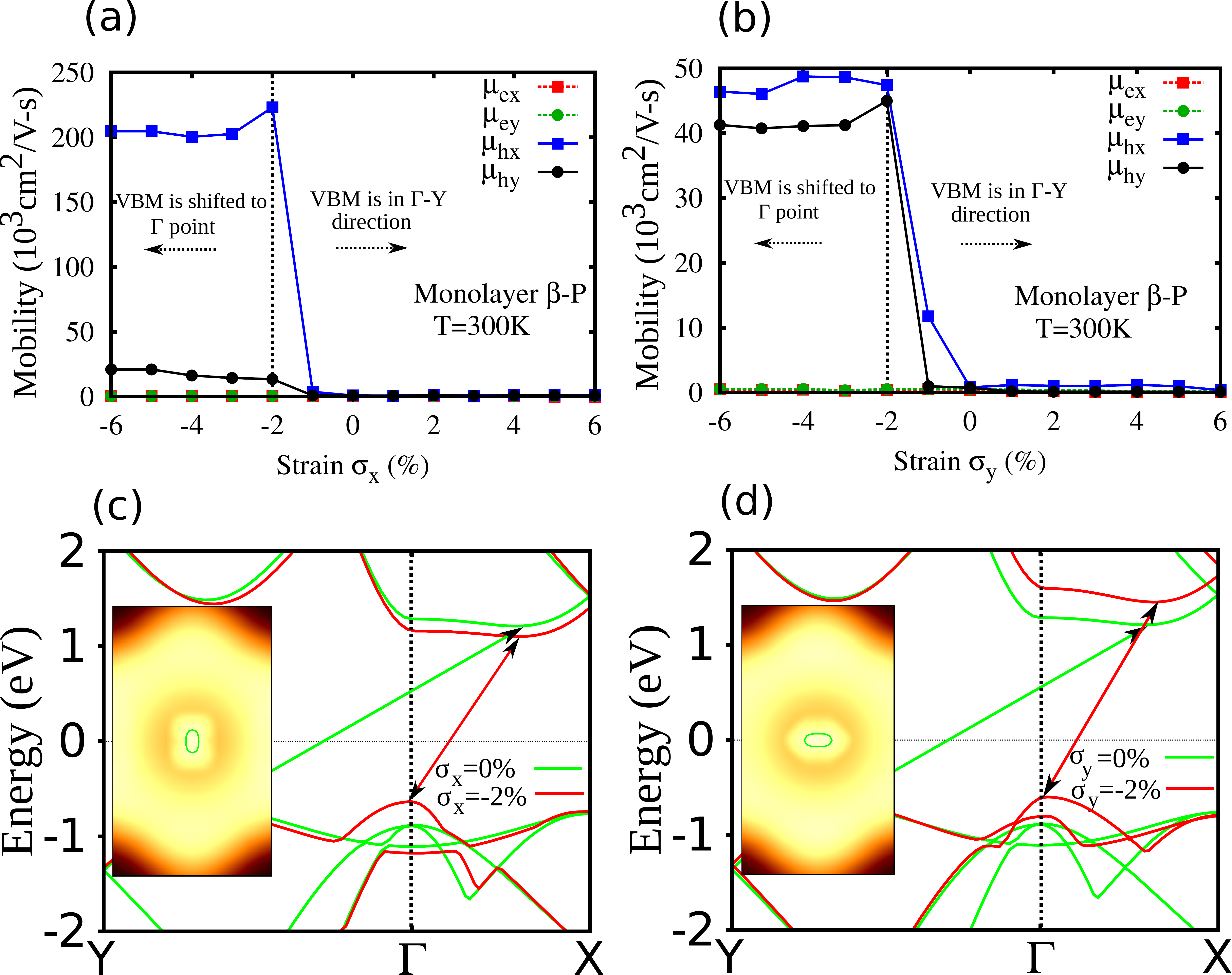}
\caption{Panels (a) and (b) display the strain dependence of electron and hole mobility of $\beta$-P. While electron mobility does not change significantly with applied strain, hole mobility increases remarkably at $\sigma_x=\sigma_y=-2\%$, both in the armchair ($x$) and the zigzag ($y$) direction. Comparing the electronic band structure of pristine [Fig~\ref{f1}(f)] and strained [panels (c) and (d)] $\beta$-P, it is observed that the VBM is shifted to the $\Gamma$ point at 2\% compressive strain, leading to decrease of hole effective mass and increase of hole mobility. Insets in panel (c) and (d) show the 2D plots of the valence band energy over the entire Brillouin zone, for $\beta$-P at $\sigma_x=-2\%$ and $\sigma_y=-2\%$.}
\label{f4}
\end{figure}

\subsection{Mobility of monolayer $\beta$-P}
Monolayer blue phosphorus or $\beta$-P is an indirect bandgap semiconductor and has the highest bandgap among all the monolayer allotropes.  Interestingly, the deformation potentials of electrons are found to be significantly higher (6 to 7 times) than that of holes; and thus the scattering probability of negative charge carries are nearly 40 times larger than that of positive charge carriers, resulting a higher mobility of hole than electron, irrespective of the measurement directions. With regards to the elastic constant, we find pristine $\beta$-P to be the least anisotropic among all the allotropes, with nearly equal $C_\alpha$ in the armchair and zigzag direction [see Table~\ref{t2}]. However, effective mass of electron (hole) is found to be 6.9 (4.5) times larger in the armchair, as compared to the zigzag direction [see Table~\ref{t2}]. On the contrary, deformation potential of electron (hole) is 2.4 (2.2) times higher in the zigzag direction and as a result, scattering probability of electron (hole) is 5.7 (4.8) times higher along the zigzag, than that of the armchair axis. Since anisotropy of effective mass and charge carrier scattering probability has opposite trend, they nearly balance each other [see Eq.~\ref{eqm}], such that the ratio  ${\mu_x}/{\mu_y}$ is equal to 0.8 and 1.08 for electrons and holes, respectively; the minimum anisotropy among the four allotropes [see Table~\ref{t2}].

When $\beta$-P is subjected to the compressive strain, hole mobility increases remarkably; by a factor of $\approx$ 250 in the armchair direction when $\sigma_x\leq -2\% $ and by a factor of $\approx$ 50 in both the directions when $\sigma_y\leq -2\% $ [see Fig.~\ref{f4} (a) and (b)]. Comparing the electronic band structure of pristine [Fig.~\ref{f1}(f)] and strained $\beta$-P [Fig.~\ref{f4}(c) and (d)], we find that the VBM is shifted to the $\Gamma$ point at 2\% compressive strain, leading to decrease of hole effective mass and increase of hole mobility. Note that, according to the 2D plot of valence band energy shown in the inset of Fig.~\ref{f4}(d), the hole effective mass is smaller along the $y$ direction; but since scattering probability is higher in this particular direction only, the resulting value of hole mobility is marginally smaller along the $y$, than that of $x$ axis. 

\begin{figure}
\includegraphics[width=\linewidth]{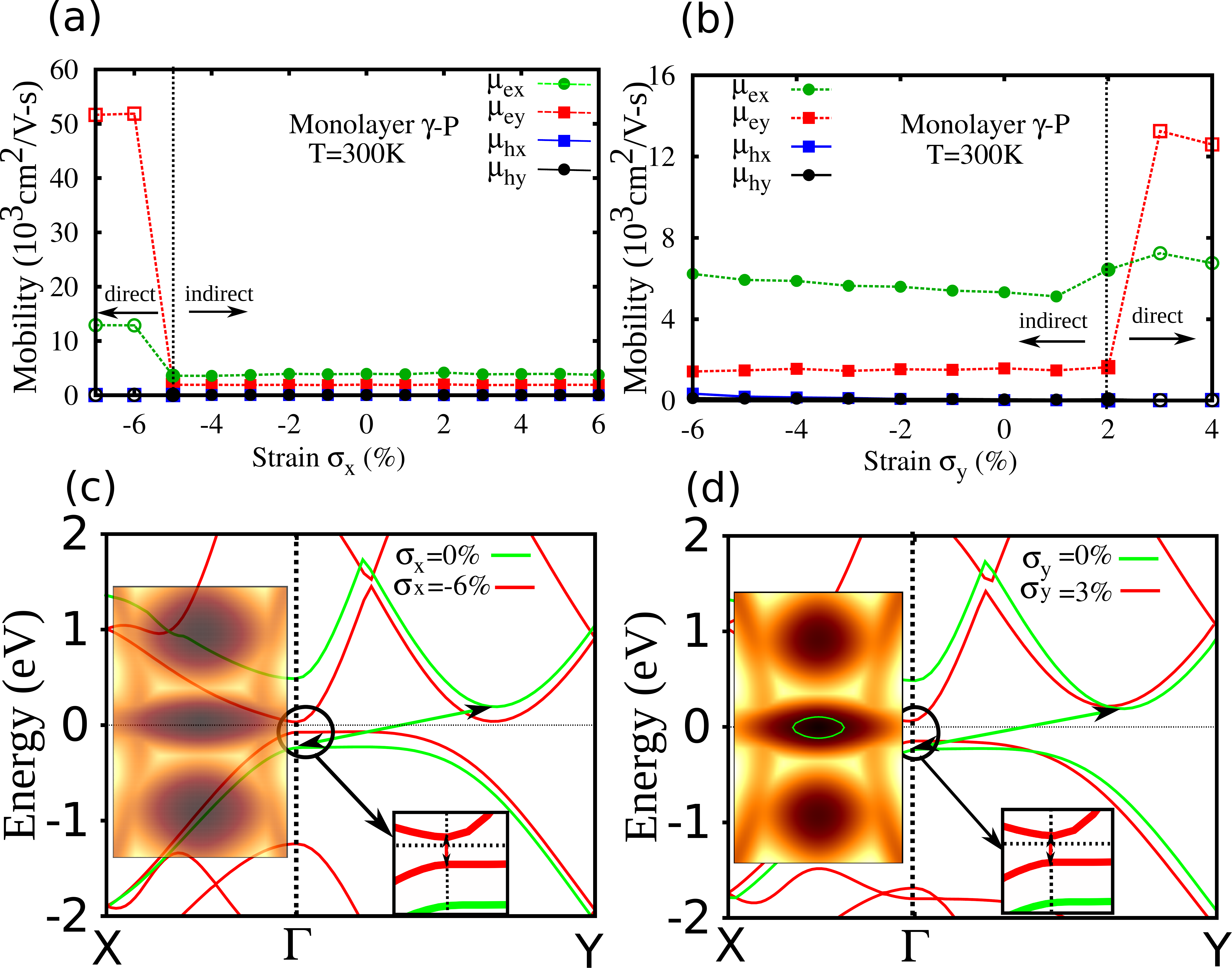}
\caption{Panels (a) and (b) display the strain dependence of electron and hole mobility of $\gamma$-P. A sharp increase of electron mobility is observed at $\sigma_x=-6\%$ and $\sigma_y=3\%$, accompanied by a reversal of preferred conduction direction from armchair ($x$) to zigzag ($y$) axis. Comparing the electronic band structure of pristine [see Fig.~\ref{f1}(g)] and strained $\gamma$-P [panels (c) and (d)], a direct to indirect bandgap transition is observed at $\sigma_x=3\%$ and $\sigma_y=-6\%$, as the CBM is relocated to the $\Gamma$ point under applied strain. The resulting decrease of the effective mass causes the mobilities to increase significantly. Insets in panel (c) and (d) show the 2D plots of the conduction band energy over the entire Brillouin zone, for $\gamma$-P at $\sigma_x=-6\%$ and $\sigma_y=3\%$.}
\label{f5}
\end{figure}

\subsection{Mobility of monolayer $\gamma$-P}
This is another indirect bandgap allotrope of monolayer phosphorus, which has a unique electronic band structure near the VBM (at the $\Gamma$ point), with a relatively flat ``Schr\"odinger like'' quadratic band dispersion along the $\Gamma$-Y (zigzag) direction, in complete contrast to a ``Dirac like'' linear band dispersion along the $\Gamma$-X (armchair) direction [see Fig~\ref{f1}(g)]. As a result, the hole effective mass is 18.9 times higher along the zigzag ($y$), than that of the armchair ($x$) axis [see Table~\ref{t2}]. Despite having such a huge effective mass anisotropy along the two principal axes, hole mobility is found to be only 4 times higher in the armchair direction, as compared to the zigzag axis [see Table~\ref{t2}]. This is because of relatively larger deformation potential in the $x$ direction, making the scattering probability to be 4.9 times higher than it's value along the $y$ axis [see Table~\ref{t2}]. As compared to the positive charge carriers, electrons have nearly two orders of magnitude higher mobility (irrespective of the measurement direction) due to a combined effect of relatively low effective mass, as well as low scattering probability. Armchair is found to be the preferred conduction direction, with a 3.5 times higher value of electron mobility than that in the zigzag direction. This is due to the combined effect of relatively low effective mass and low scattering probability (resulting from low deformation potential and high elastic constant) of electrons along the $x$ axis.

Under external strain, sharp change of electron mobilities are observed at $\sigma_x\leq -6\%$ (33 times enhancement of $\mu_{ey}$) and $\sigma_y\geq 3\%$ (10 times enhancement of $\mu_{ey}$), accompanied by a reversal of preferred direction of conduction from armchair ($x$) to zigzag ($y$) axis [see Fig.~\ref{f5}(a) and (b)]. Comparing the electronic band structures of pristine [see Fig.~\ref{f1}(k)] vs. strained $\gamma$-P [see Fig.~\ref{f5}(c) and (d)], it is observed that an indirect to direct bandgap transition takes place at $\sigma_x\leq -6\%$ and  $\sigma_y\geq 3\%$ strain, as the CBM is relocated to the $\Gamma$ point, while position of the VBM remains unchanged. As illustrated in the 2D plots of conduction band energy in the inset of Fig.~\ref{f5}(c) [$\sigma_x\leq -6\%$] and Fig.~\ref{f5}(d) [$\sigma_y\geq 3\%$], evidently the effective mass along the zigzag ($y$) axis is less, which is responsible for the electron mobility enhancement of $\gamma$-P under strain. 

\subsection{Mobility of monolayer $\delta$-P}
Except for $\alpha$-P, this is the only other direct bandgap allotrope of monolayer phosphorus. The effective mass anisotropy [$m_x^*/m_y^*$] for electron (hole) is found to be approximately $ 2.3 (2.8)$ [see Table~\ref{t2}]. Although the elastic constant along the $y$ (zigzag) direction is $\approx 1.7$ times larger, but the deformation potential of electron is $\approx 1.3 (1.2)$ times higher for electrons (holes) in the $x$ direction, which makes the scattering probability along the latter direction to be $\approx 3 (2.5)$ times more for the electrons (holes)  [see Table~\ref{t2}]. Because of smaller effective mass, as well as lower scattering probability, mobility in the $y$ direction is $\approx 6.8 (7.2)$ times higher for electrons (holes) as compared to the corresponding values in the $x$ direction [see Table~\ref{t2}]. Notably, among all the allotropes, $\delta$-P has the highest value of intrinsic mobility for both electron and hole in the $y$ direction.

\begin{figure}
\includegraphics[width=\linewidth]{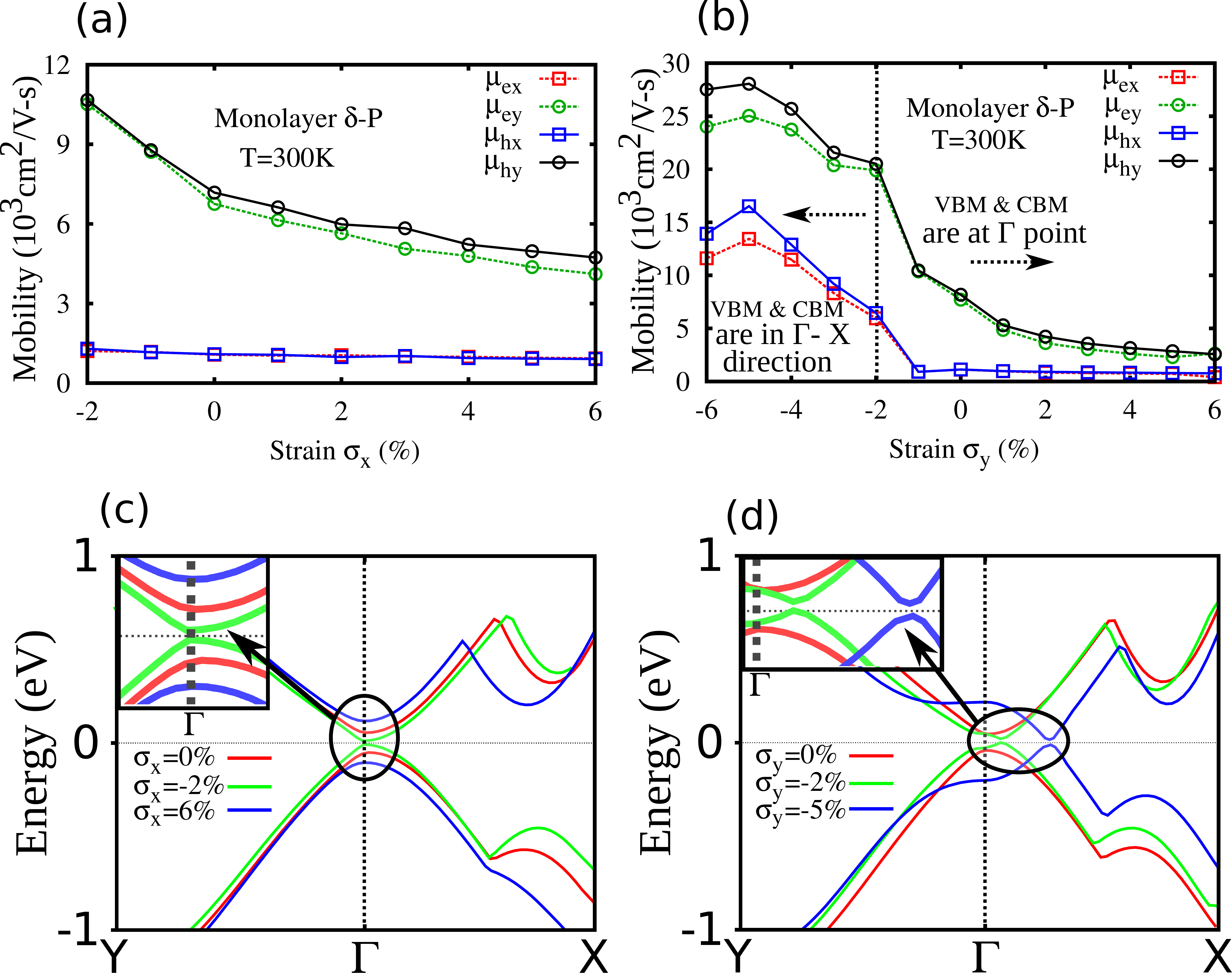}
\caption{Panels (a) and (b) display the strain dependence of electron and hole mobility of $\delta$-P. When strain is applied in the $x$ direction, $\mu_{ey}$ and $\mu_{hy}$ increases (decreases) for compressive (tensile) strain, while $\mu_{ex}$ and $\mu_{hx}$ remains unaffected. On the other hand, when compressive strain is applied in the $y$ direction, electron and hole mobilities increase in all the directions. But, if $\sigma_y$ is tensile, only $\mu_{ey}$ and $\mu_{hy}$ decreases, while $\mu_{ex}$ and $\mu_{hx}$ remains unchanged. A comparison among electronic band structures of pristine and strained $\delta$-P are shown in panels (c) and (d).}
\label{f6}
\end{figure}

When monolayer $\delta$-P is subjected to strain in the $x$ direction, mobilities of the charge carriers along the preferred conduction directions, i.e.  $\mu_{ey}$ and $\mu_{hy}$, is found to increase (decrease) for compressive (tensile) strain, while no significant change is observed in the $x$ direction [see Fig~\ref{f6}(a)]. As shown in Fig.~{\ref{f6}(a)}, the charge carrier mobilities along the zigzag axis are enhanced by a factor of $\approx 2$ in a gradual manner, as the applied strain ($\sigma_x$) is varied from 6\% tensile to 2\% compressive. On the other hand, when strain is applied in the $y$ direction, charge carrier mobilities are found to vary over a larger range, particularly in the preferred ($y$) direction of conduction [see Fig~\ref{f6}(b)]. Similar to the case of $\sigma_x$, both electron and hole mobility increases (decreases) when $\sigma_y$ is compressive (tensile). The net enhancement of $\mu_{ey}$ and $\mu_{hy}$ is found to be $\approx$ 8 to 9 times, as $\sigma_y$ is changed from 6\% tensile to 6\% compressive, with a relatively sharp increase at $\sigma_y=-2\%$ [see Fig~\ref{f6}(b)]. This is associated with a shift of both VBM and CBM from the $\Gamma$ point [see Fig~\ref{f6}(d)] and related decrease of the effective mass. Transverse to the preferred conduction direction, charge carrier mobilities (i.e., $\mu_{ex}$ and $\mu_{hx}$) show similar trend when $\sigma_y$ is compressive [see Fig~\ref{f6}(b)], with a net enhancement of electron and hole mobility by a factor of $\approx$ 4 to 5, than their intrinsic values. However, $\mu_{ex}$ and $\mu_{hx}$ do not change significantly when $\sigma_y$ is tensile [see Fig~\ref{f6}(b)].

\begin{table}[t]
\caption{Weight of the $3s$ and $3p$ orbitals of P atom at the band edges for pristine and strained allotropes (values given in the parentheses) of phosphorene. Note that, only those cases are highlighted where the band edges shfit from their original position (at zero strain) to some other location in the first Brillouin zone. Anything below 10\% weight is ignored and the corresponding value is set to zero.} 
\label{t3}
\centering
\begin{tabular}{cccccc}
\hline \hline
 & Band & $s$ & $p_x$ & $p_y$ & $p_z$\\
\hline
\hline
$\alpha$-P & VBM & 0.00  & 0.00  & 0.00 & 0.82 \\
           & \textbf{CBM} & 0.15 (0.31) & 0.21 (0.50) & 0.00 (0.00) & 0.56 (0.00)\\
\hline
$\beta$-P & \textbf{VBM} & 0.13 (0.00) & 0.00 (0.45) & 0.00 (0.45) & 0.83 (0.00)\\
          & CBM & 0.14 & 0.21  & 0.21  & 0.31 \\
\hline
$\gamma$-P & VBM & 0.12 & 0.00  & 0.00  & 0.86 \\
           & \textbf{CBM} & 0.00 (0.28) & 0.60 (0.00) & 0.00 (0.00) & 0.22 (0.54)\\
\hline
$\delta$-P & VBM & 0.13  & 0.00  & 0.00  & 0.86 \\
           & CBM & 0.14  & 0.00  & 0.00  & 0.67 \\
\hline   
\end{tabular}
\end{table}

Before concluding, let us point out the origin of the observed charge carrier mobility tuning of 2D phosphorus allotropes under strain. In all the allotropes, $3s$ and $3p$ orbitals of P hybridize to form four $sp^3$ orbitals. While each P atom (having five valence electrons) bonds with three neighboring atoms, the fourth orbital is occupied by a lone pair. By projecting the band states on atomic orbitals, it is possible to determine the weight of different orbitals in each energy band and the most vital information lies in the composition of the band edges. Weight of the $3s$ and $3p$ orbitals of P atom at the valence and conduction band edges are reported in Table~\ref{t3}, where the numbers given in the parentheses correspond to the values obtained in the strained state of the allotropes. Clearly, every sharp changes of carrier mobility associated with flipping of effective mass anisotropy is related to the shift of the VBM or CBM from it's original location (at zero strain) to some other $k$ point in the Brillouin zone [compare Table~\ref{t3} with Fig.~\ref{f3}, Fig.~\ref{f4} and Fig.~\ref{f5}].

All the calculations presented here are for the non-degenerate case. 
In case of 2D materials like graphene, MoS$_2$ and black-P, it has been shown (based on experiments and first-principles calculations) that the electron-phonon coupling generally increases linearly with electron doping, depending on the symmetry of the material. \cite{10.1038/nnano.2008.67, PhysRevB.85.161403, 2053-1583-3-1-015008}
Enhanced electron-phonon coupling means increased phonon scattering and reduced mobility. We expect similar effects to play out in all four phosphorus allotropes.

\section{Conclusion}
\label{con}
Using DFT based \textit{ab initio} calculations, we have investigated the effect of strain on electronic transport properties of the four atomically thin phosphorus allotropes, $\alpha$-P (monolayer of black phosphorus), $\beta$-P (monolayer of blue phosphorus), $\gamma$-P and $\delta$-P. We start with the pristine (zero strain) allotropes and among them, $\alpha$, $\gamma$ and $\delta$ are found to be highly anisotropic, with 4-8 times difference between the mobilities calculated along the armchair ($x$) and zigzag ($y$) direction. Other than relatively small hole mobility found in $\gamma$-P, rest of the allotropes have moderate to very high charge carrier mobility, ranging from $\approx$ 200--8000 cm$^2$V$^{-1}$s$^{-1}$. When the allotropes are subjected to external uniaxial strain in either zigzag or armchair direction, other than bandgap magnitude modification observed in all the allotropes, we also find indirect to direct bandgap (in $\gamma$-P) transition and complete closure of the bandgap (in $\gamma$ and $\delta$-P). 

In addition to this, we find that the acoustic phonon limited charge carrier mobility can be enhanced significantly (typically by a factor of $\approx 5-10$) in all the allotropes by applying external strain and as high as nearly 250 (30) times increase of hole (electron) mobility along the armchair (zigzag) direction is observed in case of $\beta$-P ($\gamma$-P) under external strain. Interestingly, in case of $\alpha$-P and $\gamma$-P, it is found that the applied strain can rotate the transport direction of electron by 90$\degree$.

\subsection*{Acknowledgements}
We acknowledge funding from DST INSPIRE scheme, SERB Fast Track Scheme for Young Scientist (SB/FTP/ETA-0036/2014), and DST Nanomission project. We also thank computer center IIT Kanpur for providing HPC facility. 
\bibliography{strain_mobility}
\end{document}